\documentclass[conference]{IEEEtran}
\IEEEoverridecommandlockouts
\usepackage{cite}
\usepackage{amsmath,amssymb,amsfonts}
\usepackage{array}
\usepackage{blkarray} 
\usepackage{makecell} 
\usepackage{algorithmic}
\usepackage{graphicx}
\usepackage[caption=false]{subfig}
\usepackage{textcomp}
\usepackage{xcolor}
\def\BibTeX{{\rm B\kern-.05em{\sc i\kern-.025em b}\kern-.08em
    T\kern-.1667em\lower.7ex\hbox{E}\kern-.125emX}}

\begin{document}

\title{Quantum Routing for Emerging Quantum Networks
}

\author{
\IEEEauthorblockN{Wenbo Shi and Robert Malaney}
\IEEEauthorblockA{\textit{School of Electrical Engineering \& Telecommunications} \\
\textit{The University of New South Wales},
Sydney, NSW, Australia. \\
}
}

\maketitle

\begin{abstract}
Quantum routing, the entanglement of an input quantum signal over multiple output paths, will be an important aspect of future quantum networks. Implementation of such routing in emerging quantum networks via the noisy quantum devices currently under development is a distinct possibility. Quantum error correction, suitable for the arbitrary noisy quantum channels experienced in the routing process, will be required. In this work, we design a combined circuit for quantum routing and quantum error correction, and carry out the first implementation of such a circuit on a noisy real-world quantum device. Under the assumption of statistical knowledge on the channel, we experimentally verify the quantum nature of the error-corrected quantum routing by determining the path-entanglement through quantum state tomography, measuring also its probability of success. The quantum error correction deployed is identified as successful in terms of improving the routing. Our experiments validate, for the first time, that error-corrected quantum routing in near-term noisy quantum-computing devices is feasible, and our detailed results provide a quantum-routing benchmark for all near-term quantum hardware.
\end{abstract}


\section{Introduction}


A quantum router, an important element of emerging quantum networks, can transmit a quantum signal from a singular input path to a coherent superposition of multiple output paths \cite{Lemr2013Resource, Bartkiewicz2014Using, yuan2015experimental, Bartkiewicz2018implementation, behera2019designing, singh2020designing}.
This key entanglement feature of a quantum router offers remarkable opportunities compared to classical routers \cite{yuan2015experimental}. 
Beyond its unique routing functionality, a quantum router also provides the only known technique that enables quantum random-access memory (a memory-access technique that  allows queries in superposition), as well as for the remote creation of a superposition of states drawn from a distributed classical memory\cite{giovannetti2008quantum, asaka2021quantum}. 
Quantum random-access memory, and therefore quantum routing,  also has implications for quantum machine learning \cite{rebentrost2014quantum, lloyd2014quantum}.

In near-term quantum networks, quantum routers can be be deployed using  Noisy Intermediate-Scale Quantum (NISQ) devices. In principle, such noisy devices can create, receive, transmit, and route qubits over quantum channels. 
They can be manufactured with various qubit types: e.g. superconducting, trapped ions, photonic, or silicon-based qubits \cite{huang2020superconducting, lekitsch2017blueprint, rudolph2017optimistic, yang2020operation}. 
Of particular interest to the wider community are the superconducting quantum computers operating at near absolute-zero temperature developed by IBM.
Presently, these NISQ devices can be considered as  first-generation quantum computers  - made available to the wider community through a cloud platform called the IBM Quantum Experience (IBM Q) \cite{ibmq}. 
This allows us to test directly the performance of quantum routing on real device of the type that may be deployed in a future network
 
However, currently no quantum router that can combat the noise channels inherent to NISQ devices has been experimentally tested.
Noisy quantum channels within the NISQ device introduce unwanted errors, largely through the entanglement between the information qubit and the environment \cite{roffe2019quantum, jahn2021holographic}.
This unwanted entanglement causes the leakage  from the defined two-level qubit space into a larger Hilbert space \cite{lu2008experimental}.
An example of one such noisy quantum channel is the amplitude damping channel, which describes the energy dissipation effects  from a quantum system, e.g. \cite{yang2010arbitrated}.
A plethora of quantum error correction protocols have been proposed to eliminate the errors caused by noisy quantum channels, e.g. \cite{al2011reversing, fletcher2008channel, grassl2018quantum, gong2022experimental}. 
In this work, we develop, and for the first time, experimentally test a combined quantum-error-corrected quantum router in a NISQ device. 
\emph{Although generic quantum error correction on current NISQ devices is not plausible, a main contribution of this work is to show  that such error correction is possible if some limited statistical information on the noise channel is known.}


All contributions of this paper can be summarized as follows.
We design a novel quantum circuit for quantum-error-corrected quantum routing, based on noisy superconducting qubits. Our circuit is built on the assumption that the noisy channel is of known characteristics, and statistical information on its key parameter is \emph{a priori} known.
The quantum circuit is then experimentally executed on a seven-qubit  NISQ device, the \textit{ibmq\_jakarta},  accessed through the IBM Q.
We verify the quantum nature of the quantum router by identifying the generation of the path-entanglement via quantum state tomography.
 The importance of the quantum error correction in regard to the routing is quantified.

The rest of this paper is organized as follows.
In Section \ref{Principle}, we introduce the basic principle of the quantum router protocol, the noisy quantum channel, and the error correction protocol.
Section \ref{Results} reports the experimental setup, the designed quantum circuit, and the results of the experiments executed on the IBM Q platform,
and Section \ref{Conclusions} concludes our work.

\section{Quantum Routing} \label{Principle}
	\subsection{Noiseless quantum routing}


\begin{figure}[tb]
\centering
\includegraphics[width=0.9\linewidth]{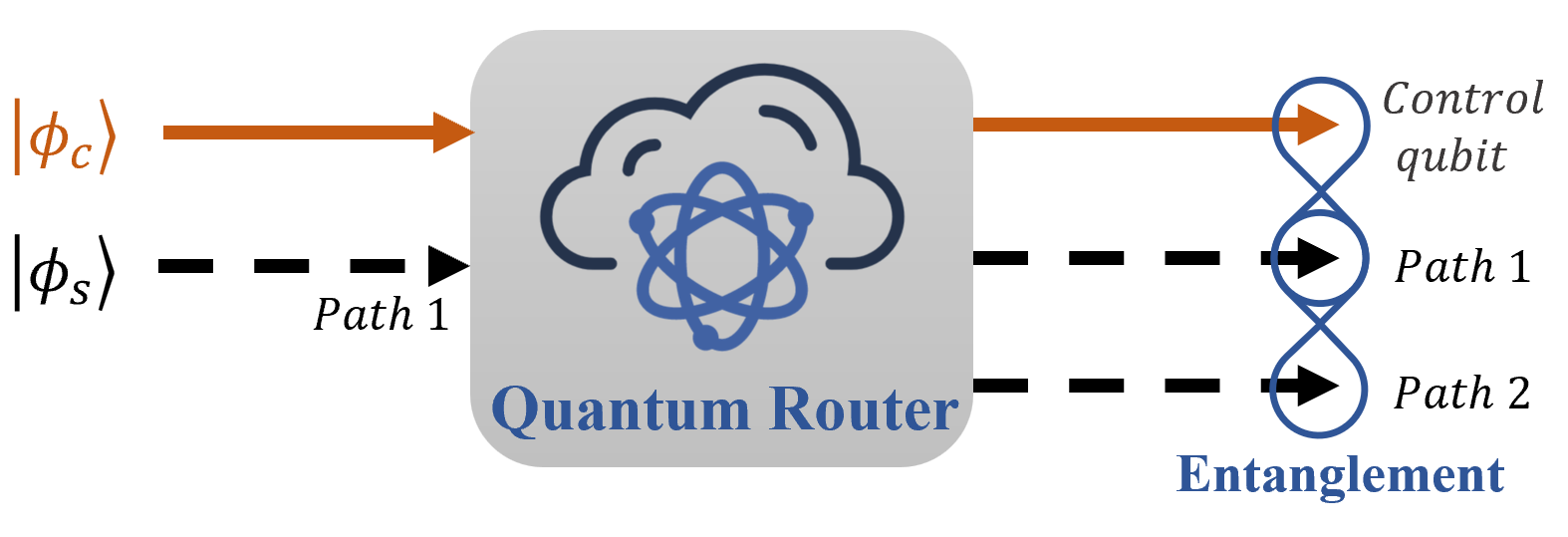}
\caption{Schematic diagram illustrating the principle of a quantum router.
The control qubit $|\phi_c\rangle$ stores the control information to direct the path of the signal qubit $|\phi_s\rangle$, which is received by the quantum router at path $1$.
The output of the quantum router is an entanglement between the control qubit and signal paths.}
\label{QRprotocol}
\end{figure}

The  fidelity of the output entangled state with the perfect routing outcome, and the total success probability of the circuit, are both considered important figures of merit - we use these to benchmark the performance of  the \textit{ibmq\_jakarta's} quantum routing.
A schematic diagram of a quantum router is shown in Fig.~\ref{QRprotocol}. 
The control qubit $|\phi_c\rangle$ contains the control information that directs the path of a signal qubit $|\phi_s\rangle$, and $|\phi_c\rangle$ can be expressed as
\begin{equation} \label{eqfirst}
|\phi_c\rangle = 
\alpha_c |0\rangle_c + \beta_c |1\rangle_c
=
\left( |0\rangle_c + |1\rangle_c\right) /\sqrt{2}
\text{,}
\end{equation}
where $\alpha_c$ and $\beta_c$ are complex numbers satisfying $|\alpha_c|^2+|\beta_c|^2 = 1$, and we define $\alpha_c = \beta_c = 1/\sqrt{2}$.
The signal information is carried by $|\phi_s\rangle$, which is received by the quantum router at path $1$, and $|\phi_s\rangle$ can be written as
\begin{equation}
|\phi_s\rangle = 
\alpha |0\rangle + \beta |1\rangle 
=
\cos \frac{\pi}{4}|0\rangle+e^{i\pi/4} \sin \frac{\pi}{4}|1\rangle
\text{,}
\end{equation}
where $\alpha$ and $\beta$ are again complex numbers satisfying a normalization constraint.
In the \textit{ibmq\_jakarta}, the quantum router requires a `blank' qubit $|\phi_n\rangle$, which contains no signal information and is initially located at path $2$. 
This is simply a function of the device's physical architecture, and its presence has no bearing on our results - it can be trivially removed if need be at the end of the process (generic quantum routing requires no such blank input state).
The input of the quantum router is $|\Phi\rangle=|\phi_c\rangle |\phi_s\rangle|\phi_n\rangle$, and the output of the quantum router $|\Phi\rangle_f$ is an entanglement between the control qubit and the two paths.
The $|\Phi\rangle_f$ takes the form
\begin{equation}
\begin{split}
|\Phi\rangle_f &=
\alpha_c |0\rangle_c |\phi_s\rangle_1|\phi_n\rangle_2 
+ \beta_c |1\rangle_c|\phi_n\rangle_1 |\phi_s\rangle_2
\text{,}
\end{split}
\end{equation}
where the subscripts $1$ and $2$ denote that the corresponding qubit is in the path $1$ and $2$, respectively.
The signal qubit is routed to the path $1$ and $2$ respectively when $|\phi_c\rangle$ is in the $|0\rangle_c$ and $|1\rangle_c$ states, and when $|\phi_c\rangle$ is a superposition state, the two paths both possess the signal qubit.

	\subsection{Noisy quantum channel}

\begin{figure}[tb]
\centering
\includegraphics[width=1\linewidth]{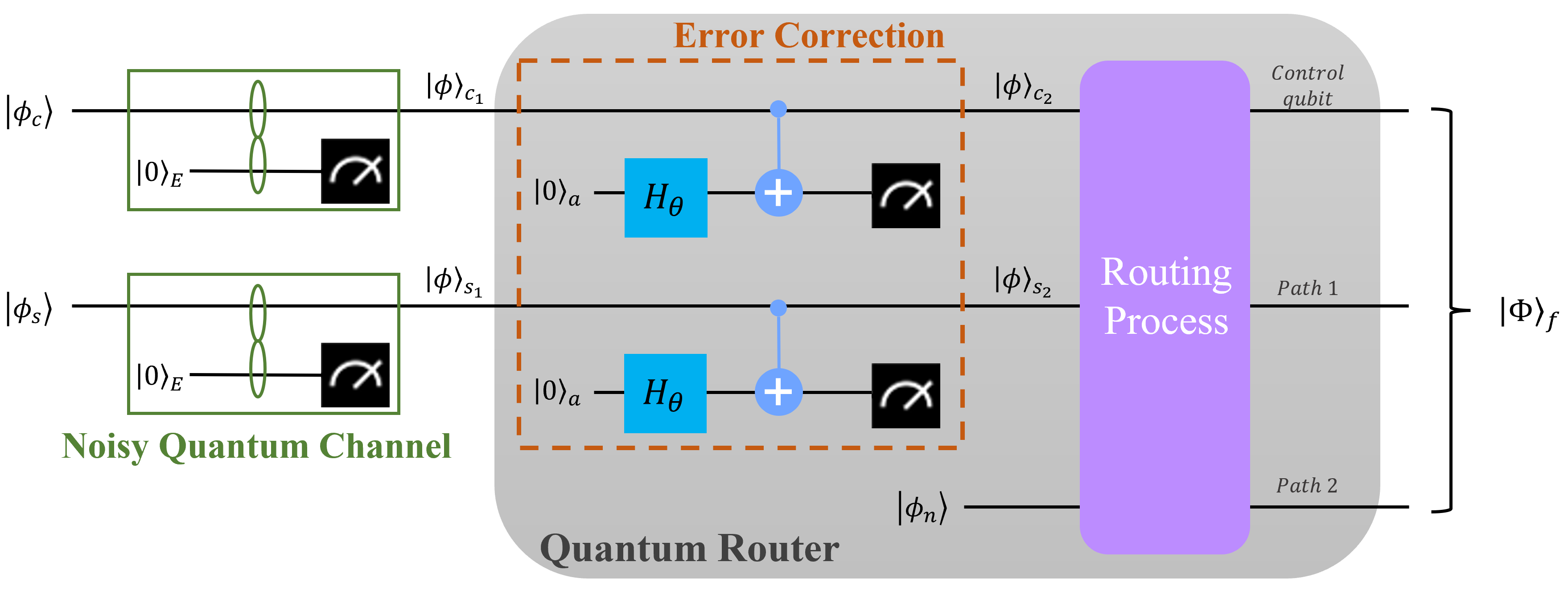}
\caption{Schematic diagram of a quantum router with noisy quantum channels and error correction.
The states $|\phi_c\rangle$ and $|\phi_s\rangle$ are prepared by some sender who sends the states through the noisy quantum channels.
Once the quantum router receives the states, error correction is conducted followed by the quantum routing process.
Note, $|0\rangle_E$ is an auxiliary qubit that simulates the environment, and $|0\rangle_a$ stands for an ancillary qubit.}
\label{QRECprotocol}
\end{figure}

While the quantum router protocol outlined  above assumes zero channel noise, we wish to consider in this work the more realistic situation where noise channels are present. 
That is, we assume the states $|\phi_c\rangle$ and $|\phi_s\rangle$ are prepared at some sender, and then passed through noisy quantum channels.
We build a parameterized noisy quantum channel that has similar characteristics to  the amplitude damping channel. 
The details of the noisy channel are not important in this work, we simply require that we have an effective method within the NISQ device to add arbitrary noise to the qubits, and that the level of that noise can be parameterized with a single parameter.

The qubits that transmitted through the noisy quantum channel can be regarded as an open system that interacts with the environment during the transmission.
We add an auxiliary qubit to simulate the environment, which starts in a pure state $|0\rangle_E$, as shown in the green boxes of Fig.~\ref{QRECprotocol}.
The amplitude damping channel models energy relaxation from an excited state to the ground state, and the evolution of $|\phi_s\rangle$ with the environment under the amplitude damping channel can be expressed as $U|\phi_s\rangle |0\rangle_E = \alpha |00_E\rangle + \beta \sqrt{\gamma}|01_E\rangle +\beta \sqrt{1-\gamma}|10_E\rangle$.
Here, $U$ is a unitary matrix which can be written as
\begin{equation}
\begin{split}
U = 
\begin{bmatrix}
1& 0&0&0\\
0& \sqrt{1-\gamma} & \sqrt{\gamma} & 0 \\
0& -\sqrt{\gamma} & \sqrt{1-\gamma} & 0 \\
0&0&0&1
\end{bmatrix}
\text{,}
\end{split}
\end{equation}
where $\gamma \in [0,1]$ is a tunable parameter.
Next, and different from amplitude damping\footnote{To realize the amplitude damping channel on a NISQ device, we would not implement the post-selections ($Z$-basis measurements) on the auxiliary qubits. Mathematically, following the unitary evolution of the combined system, the `environment' qubits would be traced out before executing the error correction.}, we make a $Z$-basis measure on the auxiliary qubit and only keep the resulting state if the measurement outcome is $|0\rangle$, and at this point, the resulting state is
\begin{equation} \label{eqmiddle}
\begin{split}
|\phi\rangle_{s_1} &=
\frac{1}{N_1}\left(\alpha |0\rangle  +\beta \sqrt{1-\gamma}|1\rangle \right)
\text{,}
\end{split}
\end{equation}
where $N_1 = \sqrt{|\alpha|^2 + |\beta|^2 (1-\gamma)}$ is a normalization factor.
This process, which delivers our required parameterized noisy channel, has a success probability of $p_1 = N_1^2$. 

To mimic the need for statistical information (only) on the channel model; after modeling the channel as described above we `forget' about our knowledge of $\gamma$. 
Rather, we assume we have only \emph{statistical} knowledge. More specifically, we assume $\gamma$ to be in the range from $0$ to $1$ with a uniform distribution. Unless otherwise specified, we use an expected mean value $\gamma_g = 0.5$ in our experiments. We will never use the known value of $\gamma$ in any of the experiments shown. Although some real-world channels could be approximated by this process, we do not claim we have truly modeled a real-world channel. We use our channel scheme to simply illustrate that when statistical information on a channel is available, quantum error correction on quantum routing within a NISQ device becomes possible.   Other, more complicated, channels will likely exist in the wide range of NISQ devices now being produced via multiple technology implementations. While we expect similar outcomes to those reported here for some of these other channels, we should be clear that the explicit results we show are specific to the statistical noise model we have assumed.


	\subsection{Error correction protocol} \label{ECP}

The correction protocol \cite{al2011reversing} we adopt is not a \emph{general} error correction scheme for an arbitrary error on a qubit, but one which assumes some \emph{a priori} knowledge of the noisy quantum channel.
Our adopted scheme applies to scenarios where some access to the entangled environment may be available \cite{superconduct} or where a weak measurement is done to detect leakage from the system  \cite{al2011reversing}, and the loss rate is known. 

We first apply a Hadamard gate $H_{\theta}$ with a parameter $\theta$ to  $|0\rangle_a$, where the Hadamard gate is,
\begin{equation}\label{Htheta}
H_{\theta} = 
\begin{bmatrix}
\cos\theta &  -\sin\theta \\
\sin\theta & \cos\theta
\end{bmatrix}
\text{.}
\end{equation}
Then, a Controlled-$X$ ($CX$) gate is performed on $|\phi\rangle_{s_1}$ and the ancillary qubit (the control and  target qubit, respectively), as illustrated in the orange dashed box of Fig.~\ref{QRECprotocol}. 
The resulting transformation can be expressed as
\begin{equation}
\begin{split}
& |\phi\rangle_{s_1} |0\rangle_a =
\frac{1}{N_1}\left(\alpha |0\rangle  +\beta \sqrt{1-\gamma}|1\rangle \right) \otimes
|0\rangle_a\\
&\xrightarrow{I \otimes H_{\theta}}
\frac{1}{N_1}\left(\alpha |0\rangle  +\beta \sqrt{1-\gamma}|1\rangle \right) \\
&\qquad \quad \otimes
\left( \cos\theta |0\rangle_a + \sin\theta|1\rangle_a \right) \\
& \xrightarrow{CX} 
\frac{1}{N_1} \Big( \alpha \cos\theta |00_a\rangle  
+\alpha \sin\theta |01_a\rangle \\
&\quad \quad 
+\beta \sqrt{1-\gamma} \cos\theta |11_a\rangle
+ \beta \sqrt{1-\gamma} \sin\theta |10_a\rangle \Big) 
\text{,}
\end{split}
\end{equation}
where $I$ is the identity matrix.
The last step of the error correction requires a post-selection method applied to the ancillary qubit. This method involves the retention of the post-selected state only when the $Z$-basis measurement result of the ancillary qubit is $|0\rangle$.
The resulting post-selected state by this process can be written,
\begin{equation} \label{eqfinal}
\begin{split}
|\phi\rangle_{s_2} &= 
\frac{1}{N_1N_2}\left(\alpha \cos\theta |0\rangle  
+\beta \sqrt{1-\gamma} \sin\theta  |1\rangle \right) 
\text{,}
\end{split}
\end{equation}
where $N_2$ is a normalization factor expressed as $N_2 =\sqrt{|\alpha \cos\theta |^2 + |\beta  \sin\theta|^2 (1-\gamma)} / N_1$.
When we set $\theta = \arctan(1/\sqrt{1-\gamma})$, giving $\cos\theta / \sin\theta = \sqrt{1-\gamma}$, $|\phi\rangle_{s_2} =|\phi_s\rangle$ with the success probability of the error correction\footnote{The process is similar when the noisy quantum channel and the error correction are applied on $|\phi_c\rangle$. Here, we only used $|\phi_s\rangle$ as an example for demonstrating the derivations.} $p_2 = N_2^2$.



\section{Experimental Results} \label{Results}
	\subsection{Experimental setup}

\begin{figure}[tb]
\centering
\includegraphics[width=0.65\linewidth]{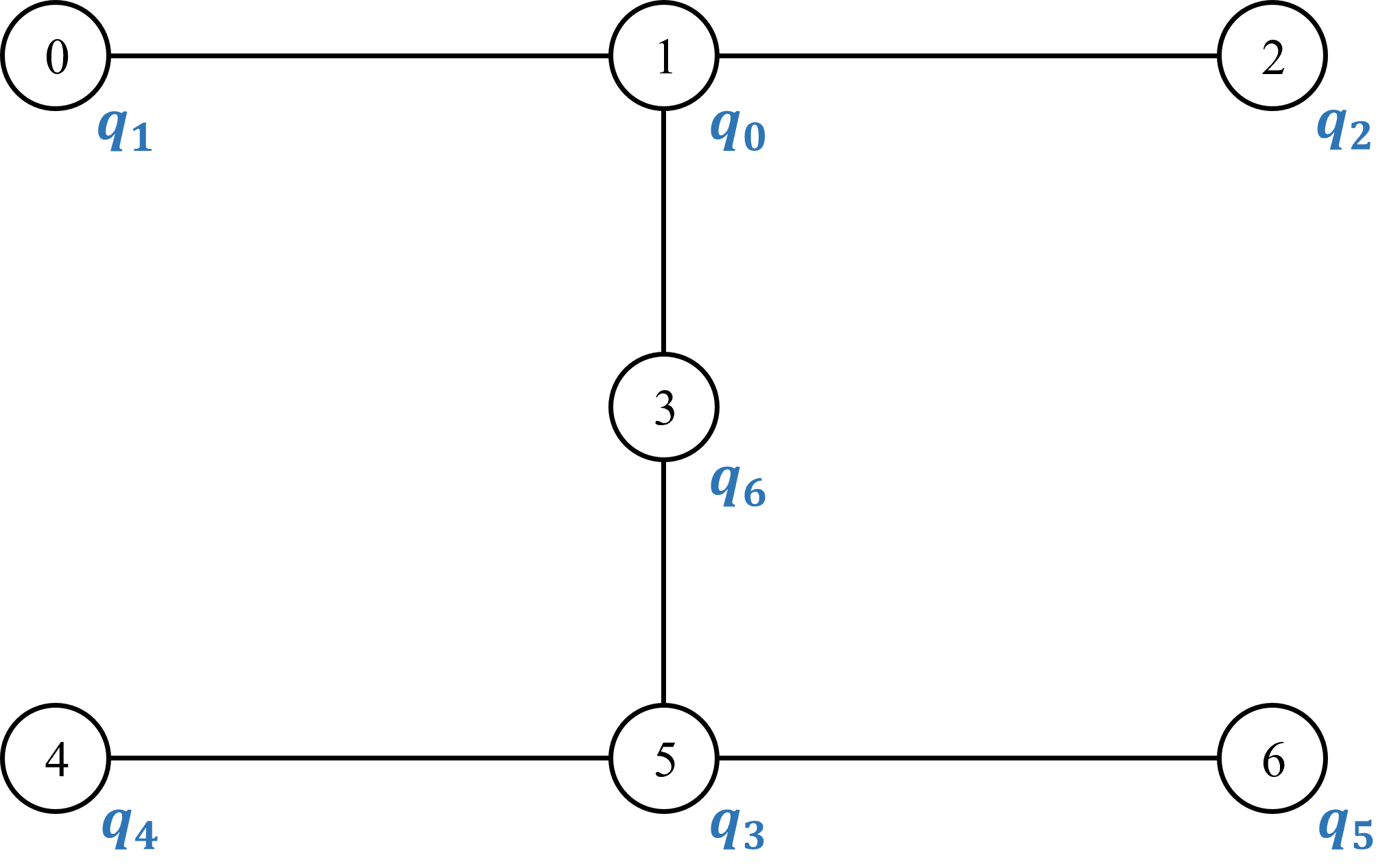}
\caption{Layout of the quantum device \textit{ibmq\_jakarta}.
The digits in the circles indicate the physical qubit number of the \textit{ibmq\_jakarta}.
The characters located at the outside of the circles correspond to the qubits labelled in Fig.~\ref{IBMCir}.
}
\label{Layout}
\end{figure}

\begin{table}[tbp]
\caption{Basis Gates of the \textit{ibmq\_jakarta}}
\begin{center}
\begin{tabular}{|c|c|}
\hline
\textbf{Basis Gates} &
\makecell*[c]{ \textbf{Matrix Representation}}  \\
\hline
\makecell*[c]{Identity gate ($I$)} 
&
\Gape[\jot]{$I$ =
$\begin{bmatrix}
1 &0  \\
0 & 1 
\end{bmatrix}$}\\
\hline
\makecell*[c]{Single-qubit rotation\\about the $Z$ axis ($Rz$)} 
&
\Gape[\jot]{$Rz(\lambda) ^{\mathrm{a}}$ =
$\begin{bmatrix}
e^{-i\frac{\lambda}{2}} &0  \\
0 & e^{i\frac{\lambda}{2}} 
\end{bmatrix}$}\\
\hline
\makecell*[c]{Single-qubit $\sqrt{X}$ gate} 
&
\Gape[\jot]{$\sqrt{X}$ =
$\dfrac{1}{2}$
$\begin{bmatrix}
1+i &1-i   \\
1-i & 1+i 
\end{bmatrix}$} \\
\hline
\makecell*[c]{Single-qubit $X$ gate} 
&
\Gape[\jot]{$X$ =
$\begin{bmatrix}
0 &1   \\
1 & 0 
\end{bmatrix}$}\\
\hline
\makecell*[c]{Two-qubit\\Controlled-$X$ gate ($CX$)} 
&
\Gape[\jot]{$CX^{\mathrm{b}}$ =
$\begin{bmatrix}
1& 0& 0 &0   \\
0& 0& 0 &1   \\
0& 0& 1 &0   \\
0& 1& 0 &0   
\end{bmatrix}$}\\
\hline
\multicolumn{2}{l}{$^{\mathrm{a}} \lambda$ is a phase term.}\\
\multicolumn{2}{l}{$^{\mathrm{b}}$ Qiskit uses little-endian order.}\\
\end{tabular}
\label{tabibm}
\end{center}
\end{table}

Our experiments are implemented on the quantum device, \textit{ibmq\_jakarta}, via the  Quantum Information Science toolKit (Qiskit) -
the  open-source software development kit for creating and running quantum circuits on the IBM Q \cite{aleksandrowicz2019qiskit}.
We use  Qiskit to design the quantum circuits which are then submitted  to the \textit{ibmq\_jakarta}, which processes and returns the  measurement results.

The \textit{ibmq\_jakarta} has seven superconducting qubits in a horizontal H-shaped geometry, as illustrated in Fig.~\ref{Layout}.
The numbers in the circles label physical qubits in the \textit{ibmq\_jakarta}, and the symbols outside the circles represent the initial layout of the qubits in the designed quantum circuit.
The \textit{ibmq\_jakarta} only supports five basis gates, namely the single-qubit gates $I$, $Rz$, $\sqrt{X}$, and $X$, and the two-qubit gate $CX$, as shown in Table.~\ref{tabibm}.
The quantum circuit we submit to the \textit{ibmq\_jakarta} is transpiled to a circuit that only includes basis gates before its execution.
We execute the quantum circuit many thousands of times for each parameter setting to ensure reliable statistics (up to 100,000 - the maximum possible number for the \textit{ibmq\_jakarta}).
Note, the measurement errors (which are distinct from the errors introduced by the noisy quantum channel) are one type of intrinsic errors of the quantum device and are reduced by the default readout error mitigation package in Qiskit's Ignis library \cite{aleksandrowicz2019qiskit}.
The main idea of this readout mitigation is to measure the qubits in every basis state and to then calculate the probabilities for all possible measurement results. 
In implementation, these probabilities are used to build a calibration matrix,  the inverse of which is applied to the experimental outcomes, eliminating the measurement errors in the ideal case.

	\subsection{The quantum circuit for error-corrected routing} \label{Qcir}

\begin{figure}[tb]
\centering
\includegraphics[width=\linewidth]{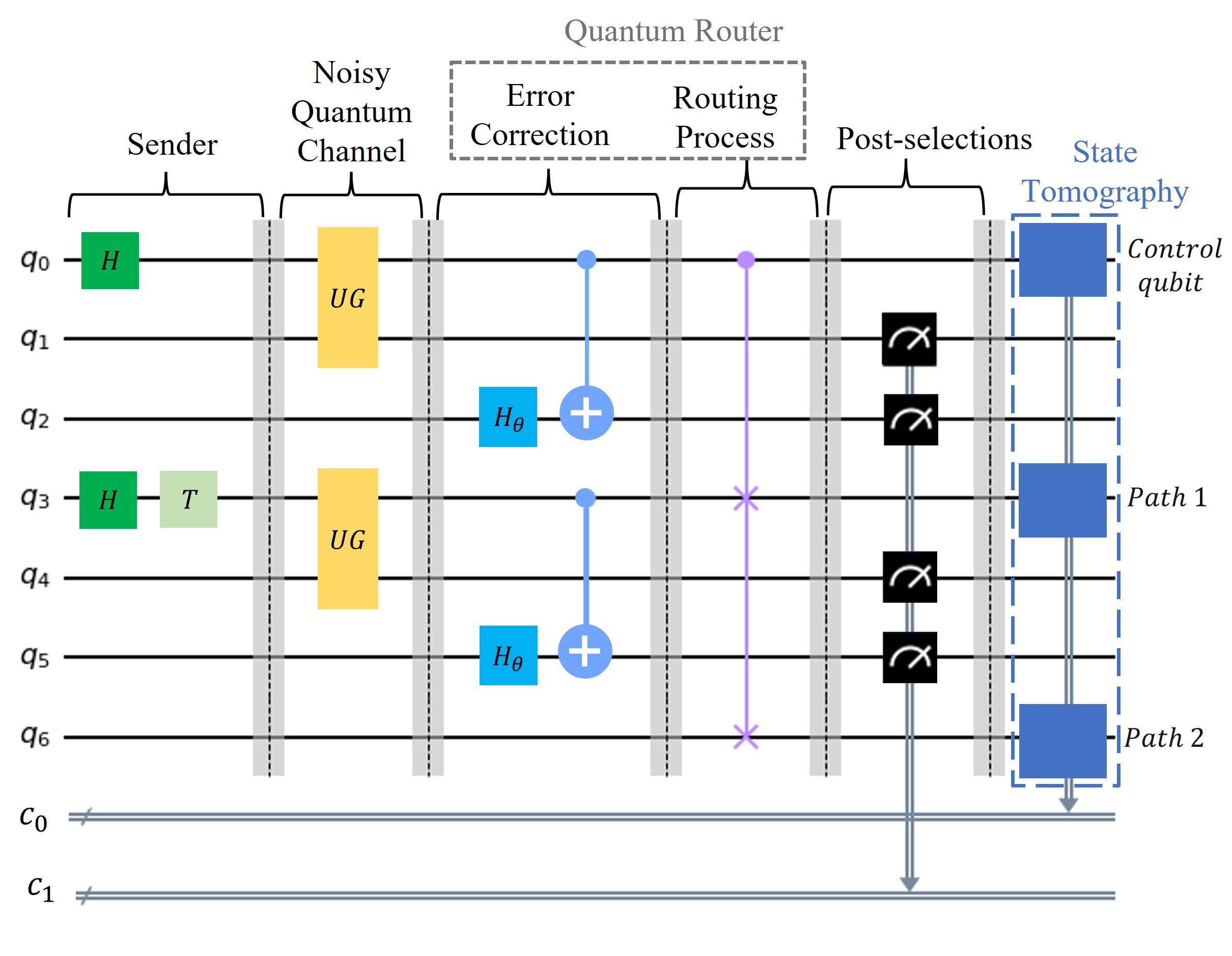}
\caption{Quantum circuit of the quantum router with state tomography. 
The qubits $q_0$, $q_3$, and $q_6$ are prepared as $|\phi_c\rangle$, $|\phi_s\rangle$, and $|\phi_n\rangle$, respectively.
$q_1$ and $q_4$ are auxiliary qubits initialized in the state $|0\rangle_E$.
$q_2$ and $q_5$ are the ancillary qubits each of which is written as $|0\rangle_a$.
$H$ represents the Hadamard gate, and $T$ the single-qubit phase gate, which induces a $\pi/4$ phase. 
$UG$ is a unitary gate represented by the unitary matrix $U$, and $H_{\theta}$ is the Hadamard gate represented by \eqref{Htheta}. 
The two-qubit gates applied to  $q_0q_2$ and $q_3q_5$ are $CX$ gates.
The three-qubit gate implemented on the qubits $q_0$, $q_3$, and $q_6$ is a controlled-swap gate, whose function is to swap the positions of  $q_3$ and $q_6$ when  $q_0$ is in the $|1\rangle$ state.
$c_0$ and $c_1$ are classical registers used for storing the measurement results in the state tomography and the post-selection, respectively.}
\label{IBMCir}
\end{figure}

The quantum circuit of the quantum router (alongside the noisy quantum channel and the error correction) is shown in Fig.~\ref{IBMCir} (note that all qubits are initialized as the $|0\rangle$ state).
The qubits $q_0$, $q_3$, and $q_6$ are prepared to $|\phi_c\rangle$, $|\phi_s\rangle$, and $|\phi_n\rangle$ states via single qubit gates.
$q_1$ and $q_4$ are two auxiliary qubits that simulate the environment (initialized in the $|0\rangle_E$ state). $q_2$ and $q_5$ stand for the ancillary qubits initialized in the $|0\rangle_a$ state.
The gates $UG$ are unitary gates derived from the unitary matrix $U$, used to parameterize the  noisy quantum channel.
As discussed earlier, to realize the noisy channel on the \textit{ibmq\_jakarta},  post-selections ($Z$-basis measurements) are performed on the two auxiliary qubits, $q_1$ and $q_4$. 
To implement the error correction, post-selections are also performed on qubits $q_2$,  and $q_5$ (again $Z$-basis measurements).
The classical registers $c_1$ contains all four measurement outcomes.
A controlled-swap gate is performed on $q_0$, $q_3$, and $q_6$ to realize the quantum routing process, which is the core part of the quantum router, and we can also consider the experiments as tests for the controlled-swap gate.
Finally, we note the classical register $c_0$ contains the measurement outcomes of the state tomography (see later discussion). 

We choose as one of our performance metrics the fidelity $F$ between $\sigma = |\Phi\rangle_f \langle\Phi|$ and $\sigma'$.
Here, $\sigma'$ is a reconstructed density matrix of the qubits $q_0$, $q_3$, and $q_6$ determined via the state tomography\footnote{The state tomography reconstructs the complete density matrix of a qubit through a series of measurements in the $X$-, $Y$-, and $Z$-basis. To reconstruct an $n$-qubit system, the state tomography requires $3^n$ measurements.}.
This fidelity $F$ is calculated as 
\begin{equation}
\begin{split}
F =  \left( \text{Tr} \sqrt{ \sqrt{\sigma} \,\, \sigma' \sqrt{\sigma}} \right) ^2
\text{,}
\end{split}
\end{equation}
where Tr represents the trace operation.

\begin{figure}[tbp]
\centering
\includegraphics[width=0.92\linewidth]{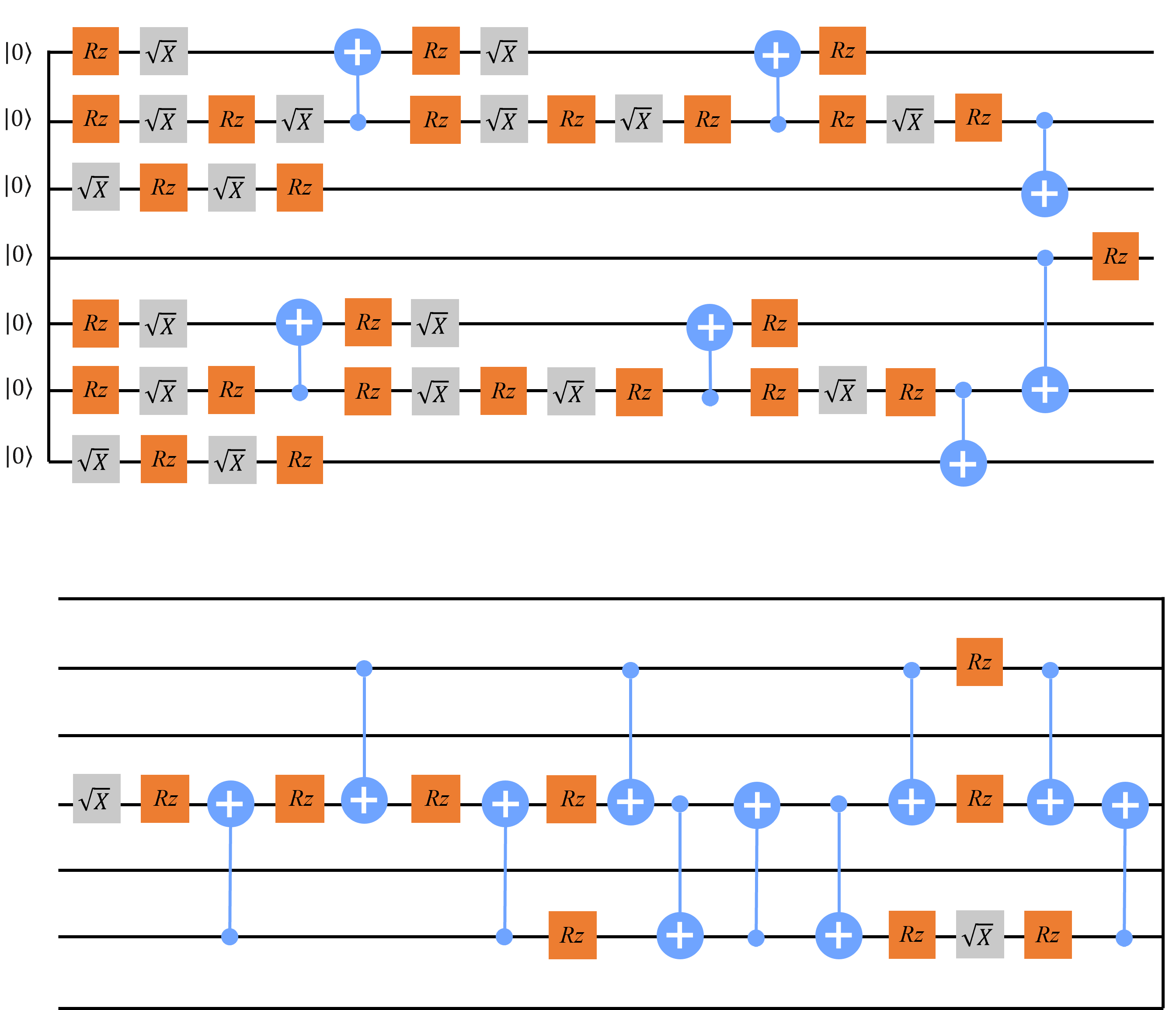}
\caption{An example of the transpiled quantum circuit of the circuit shown in Fig.~\ref{IBMCir} without the post-selection and the state tomography.
The seven qubits are the same as those shown in Fig.~\ref{Layout}, and the second row is the continuation of the first row.
Note, the qubits' positions might be changed to perform the $CX$ gate, as the two-qubit gate can only be applied to the connected physical qubits.
}
\label{TCir}
\end{figure}

As the \textit{ibmq\_jakarta} can only perform using its intrinsic basis gates, the quantum circuit is transpiled automatically (by the IBM Q) before its execution, where the transpilation process is not deterministic.
An example of the transpiled circuit is depicted in Fig.~\ref{TCir}.
Note, the qubit positions might be swapped to implement the $CX$ gate, as the two-qubit gate can only be realized on any real device between connected physical qubits.

	\subsection{Results}

\begin{figure}[tbp]
\centering
\includegraphics[width = 1\linewidth]{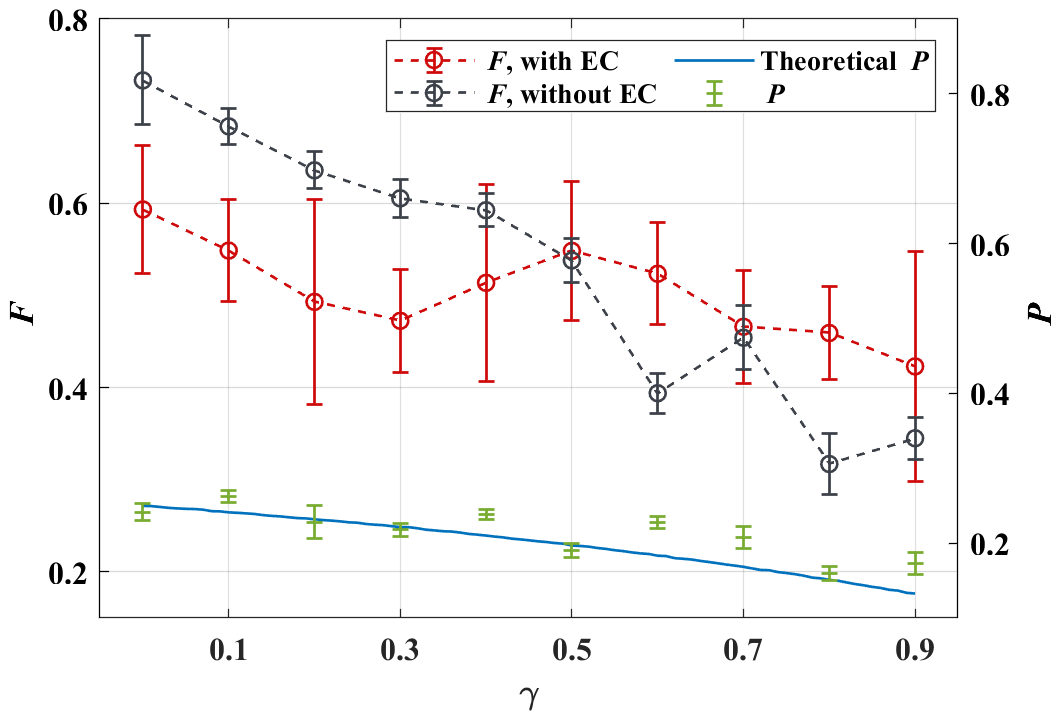}
\caption{$F$ and $P$ as a function of $\gamma$ with $\gamma_g = 0.5$ for the quantum router with and without error correction. 
Each circle and horizontal line shows the average result from ten repetitions - error bars correspond to two standard deviations assuming a Gaussian error distribution.
EC represents error correction.
}
\label{2DFr}
\end{figure}

\begin{figure}[tbp]
\centering
\includegraphics[width = 0.97\linewidth]{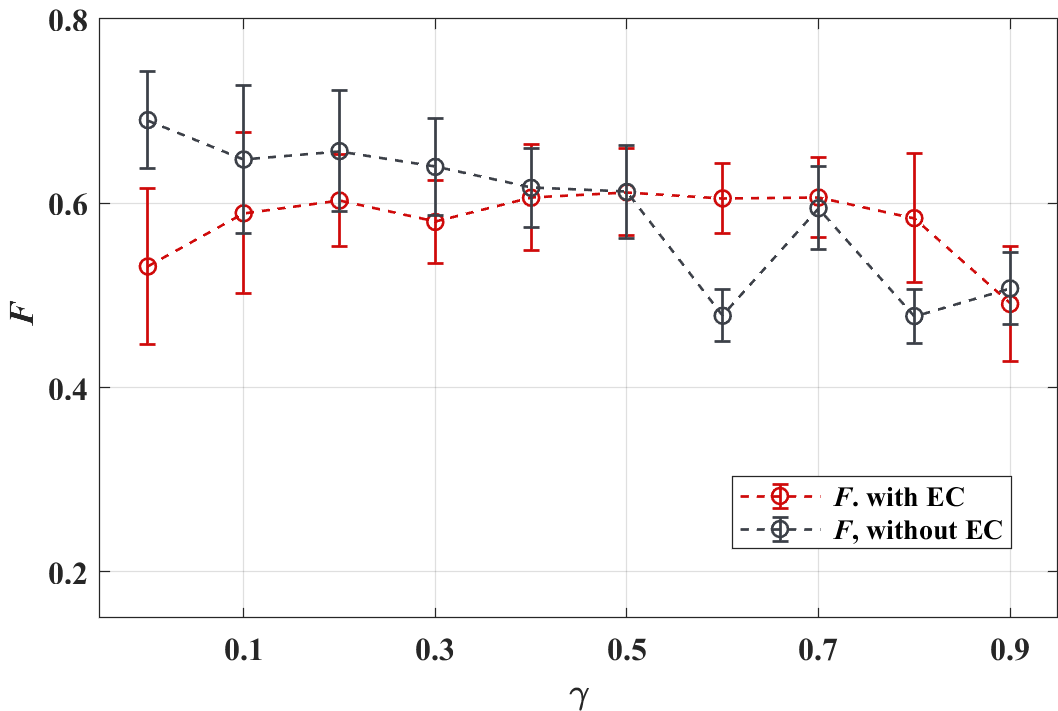}
\caption{$F$ as a function of $\gamma$ for the quantum router protocol, 
where the noisy quantum channel and error correction are only performed on $|\phi_s\rangle$.
Error bars are as described before.
}
\label{Noise}
\end{figure}

\begin{figure*}[tbp]
\centering
 \subfloat[ \label{ThR}]{ 
       \includegraphics[width=0.32\textwidth]{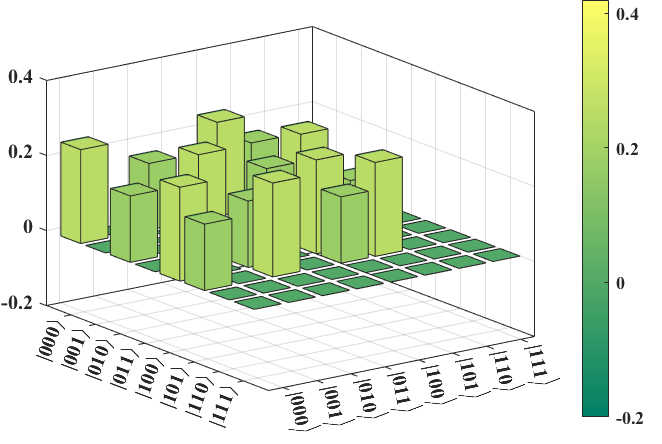}}
 \subfloat[ \label{ExRNO}]{ 
       \includegraphics[width=0.32\textwidth]{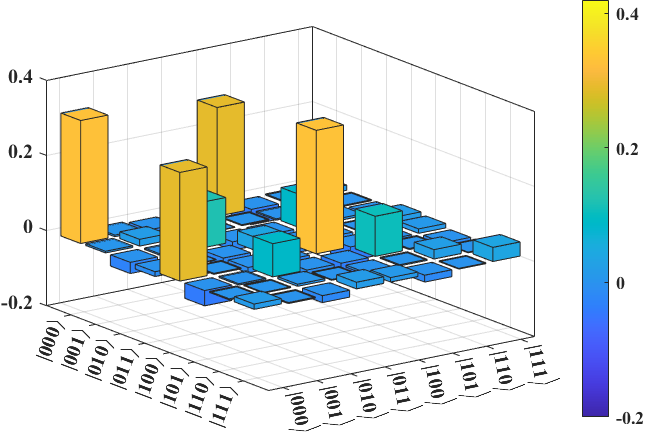}}
 \subfloat[ \label{ExR}]{ 
       \includegraphics[width=0.32\textwidth]{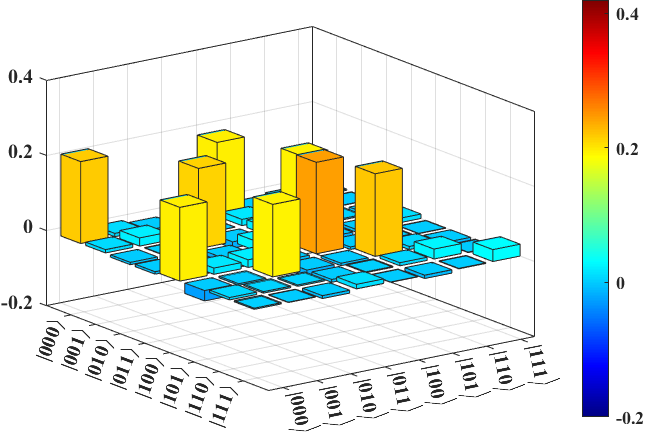}} 
\\ 
\subfloat[\label{ThI}]{ 
       \includegraphics[width=0.32\textwidth]{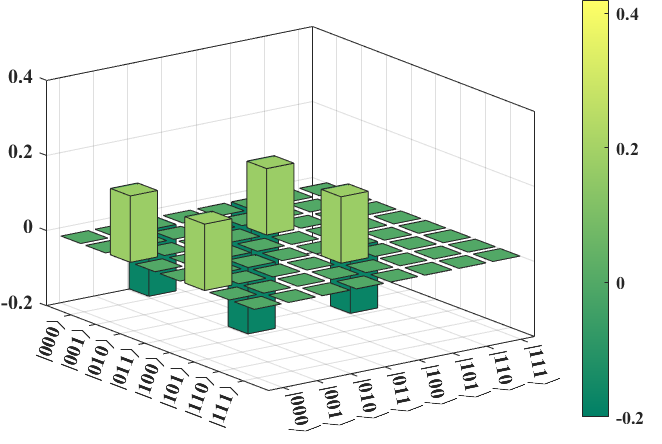}}
 \subfloat[\label{ExINO}]{ 
       \includegraphics[width=0.32\textwidth]{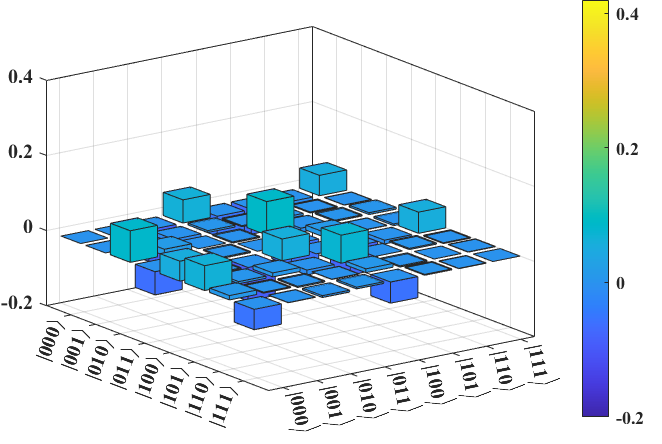}}
 \subfloat[\label{ExI}]{ 
       \includegraphics[width=0.32\textwidth]{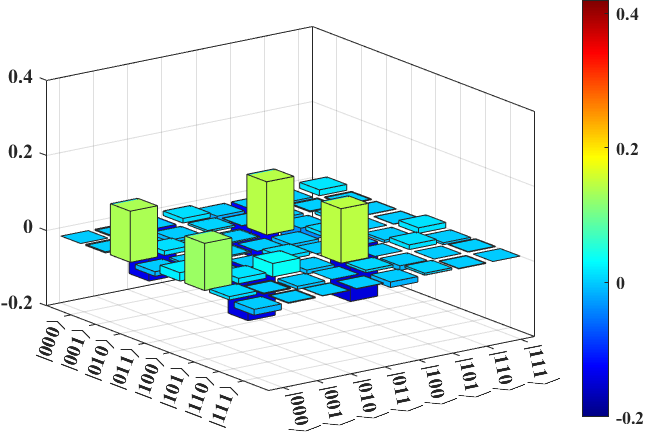}}
\caption{Theoretical and experimental density matrices of $|\Phi\rangle_f$, the entanglement generated at the output of the quantum router.
(a), (d) represent the real and imaginary parts of the theoretical density matrix, $\sigma$, respectively.
(b), (e) show the real and imaginary parts of $\sigma'$ without the error correction after the noisy quantum channel, with $\gamma=0.6$, performed on $|\phi_s\rangle$ only.
Here, $F$ between $\sigma$ and $\sigma'$ is 0.48.
(c), (f) depict the real and imaginary parts of $\sigma'$ with the noisy quantum channel and error correction applied on $|\phi_s\rangle$ ($\gamma=0.6$ and $\gamma_g = 0.5$). 
$F$ improves to 0.61 after considering the error correction.
}
\label{DM}
\end{figure*}

We see from  Fig.~\ref{2DFr} that the quantum router with the error correction is feasible when $\gamma \geq 0.5$ on the \textit{ibmq\_jakarta}. For this range of $\gamma$, the EC (error corrected) experiments are significantly above the experiments without EC - demonstrating its importance.
Note,  a baseline is the case where no noisy channel and no error correction is applied, for which the fidelity of the quantum routing is 0.85 (non-unity as a consequence of intrinsic errors within the device).
Each circle represents an outcome averaged over ten repetitions, with an error bar indicating two standard deviations from the mean. 
It is well known that with fewer quantum gates less noise is introduced to the IBM Q NISQ devices \cite{das2021design, zhukov2019quantum}.
When $\gamma < 0.5$, the noise induced by our noisy quantum channel is smaller than the noise accumulated from the quantum gates\footnote{We do note however, there remains a small probability that this situation observed at low $\gamma$ is an artifact of experimental noise as evidenced by the error bars shown.}.
This is the reason why the experimental $F$ is decreased after the error correction when $\gamma < 0.5$. 
The observed  overall trend of $F$ with $\gamma$ is as expected - the  quantum channel introduces more noise for larger $\gamma$ and the \textit{ibmq\_jakarta}  performs the error correction less efficiently as the noise increases.

Also shown (right-hand scale) in Fig.~\ref{2DFr} is the  success probability $P$ of the whole procedure (quantum-error-corrected quantum routing). 
This is given by  $P= p_2^2$, where $p_2$ is the error correction probability for a noisy qubit derived earlier (the success probability of the routing circuit is one). 
It can be  seen here that higher $\gamma$ results in lower $P$ and $F$, where the experimental $P$ is consistent with the theory.
The success probability, $P$, decreases with the increase of $\gamma$ and approaches 0 as $\gamma \rightarrow 1$, which indicates a tradeoff between $P$ and error correction. 


We also consider the scenario where the noisy quantum channel and the error correction are implemented only on one qubit, the results of which are shown in Fig.~\ref{Noise}.
Here, we see the quantum router with the error correction is better demonstrated as feasible for the whole range of $\gamma$ on the \textit{ibmq\_jakarta}. 
This phenomenon verifies again that utilizing fewer quantum gates helps improve the performance of the NISQ devices.

Finally, the quantum nature of the router is demonstrated by the entanglement generated at the output.
We verify the entanglement by reconstructing its density matrix via state tomography, the results of which are in Fig.~\ref{DM}.
The theoretical density matrix, $\sigma$, is demonstrated in Fig.~\ref{ThR} and Fig.~\ref{ThI}, and the experimental $\sigma'$ with the noisy quantum channel and error correction implemented on $|\phi_s\rangle$ only, with $\gamma= 0.6$ and $\gamma_g = 0.5$, is illustrated in Fig.~\ref{ExR} and Fig.~\ref{ExI}.
For comparison, we also demonstrated $\sigma'$ without the error correction after the noisy quantum channel performed, with $\gamma= 0.6$, on $|\phi_s\rangle$ (Fig.~\ref{ExRNO} and Fig.~\ref{ExINO}).
From the comparison of these figures, the good performance of the error-corrected quantum routing is verified - the corrected state clearly being closer in its matrix elements to the theoretical density matrix elements. 
Detailed information on the values of these elements can be seen from the range of values shown, and fidelities between the matrices determined.
It can be found that $F$ improves from $0.48$ to $0.61$ after the error correction.


\section{Conclusions} \label{Conclusions}

In this work, we designed and experimentally demonstrated a quantum router embedded with a quantum error correction scheme. 
Via quantum  state tomography, we verified the quantum nature of the router, and the impact of the error correction on the routing performance. Our results demonstrate, for the first time, that quantum routing with embedded quantum error correction is viable in near-term noisy devices - pointing the way towards the inclusion of quantum routing in emerging quantum networks. Although we have used a specifically-designed noisy channel, our work demonstrates that with use of statistical information only, quantum-error-corrected routing is viable in NISQ devices. Inclusion of quantum routing within emerging quantum networks will enhance the functionality of such networks, allow for deployment of random access memory, and provide a bridge to more enhanced quantum features. We encourage further development of the ideas presented here in the context of other noisy channels on NISQ devices.


\begin{thebibliography}{10}
\providecommand{\url}[1]{#1}
\csname url@samestyle\endcsname
\providecommand{\newblock}{\relax}
\providecommand{\bibinfo}[2]{#2}
\providecommand{\BIBentrySTDinterwordspacing}{\spaceskip=0pt\relax}
\providecommand{\BIBentryALTinterwordstretchfactor}{4}
\providecommand{\BIBentryALTinterwordspacing}{\spaceskip=\fontdimen2\font plus
\BIBentryALTinterwordstretchfactor\fontdimen3\font minus
  \fontdimen4\font\relax}
\providecommand{\BIBforeignlanguage}[2]{{%
\expandafter\ifx\csname l@#1\endcsname\relax
\typeout{** WARNING: IEEEtran.bst: No hyphenation pattern has been}%
\typeout{** loaded for the language `#1'. Using the pattern for}%
\typeout{** the default language instead.}%
\else
\language=\csname l@#1\endcsname
\fi
#2}}
\providecommand{\BIBdecl}{\relax}
\BIBdecl

\bibitem{Lemr2013Resource}
K.~Lemr \emph{et~al.}, ``Resource-efficient linear-optical quantum router,''
  \emph{Phys. Rev. A}, vol. 87, 062333, 2013.

\bibitem{Bartkiewicz2014Using}
K.~Bartkiewicz, A.~\ifmmode~\check{C}\else \v{C}\fi{}ernoch, and K.~Lemr,
  ``Using quantum routers to implement quantum message authentication and
  {Bell-state} manipulation,'' \emph{Phys. Rev. A}, vol. 90, 022335, 2014.

\bibitem{yuan2015experimental}
X.~X. Yuan, J.-J. Ma, P.-Y. Hou, X.-Y. Chang, C.~Zu, and L.-M. Duan,
  ``Experimental demonstration of a quantum router,'' \emph{Sci. Rep.}, vol.~5,
  no. 1, 12452, 2015.

\bibitem{Bartkiewicz2018implementation}
K.~Bartkiewicz \emph{et~al.}, ``Implementation of an efficient linear-optical
  quantum router,'' \emph{Sci. Rep.}, vol.~8, no. 1, 13480, 2018.

\bibitem{behera2019designing}
B.~K. Behera, T.~Reza, A.~Gupta, and P.~K. Panigrahi, ``{Designing quantum
  router in IBM quantum computer},'' \emph{Quantum Inf. Process.}, vol.~18, no.
  11, 328, 2019.

\bibitem{singh2020designing}
A.~Singh, B.~Behera, and P.~Panigrahi, ``Designing a quantum router based on
  system {Hamiltonian}: {An IBM} quantum experience,'' 2020, {doi}:
  10.13140/RG.2.2.21632.17923.

\bibitem{giovannetti2008quantum}
V.~Giovannetti, S.~Lloyd, and L.~Maccone, ``Quantum random access memory,''
  \emph{Phys. Rev. Lett.}, vol. 100, 160501, 2008.

\bibitem{asaka2021quantum}
R.~Asaka, K.~Sakai, and R.~Yahagi, ``Quantum random access memory via quantum
  walk,'' \emph{Quantum Sci. Technol.}, vol.~6, no. 3, 035004, 2021.

\bibitem{rebentrost2014quantum}
P.~Rebentrost \emph{et~al.}, ``Quantum support vector machine for big data
  classification,'' \emph{Phys. Rev. Lett.}, vol. 113, 130503, 2014.

\bibitem{lloyd2014quantum}
S.~Lloyd, M.~Mohseni, and P.~Rebentrost, ``Quantum principal component
  analysis,'' \emph{Nat. Phys.}, vol.~10, no.~9, pp. 631--633, 2014.

\bibitem{huang2020superconducting}
H.-L. Huang, D.~Wu, D.~Fan, and X.~Zhu, ``Superconducting quantum computing: A
  review,'' \emph{Sci. China Inf. Sci.}, vol.~63, no. 8, 180501, 2020.

\bibitem{lekitsch2017blueprint}
B.~Lekitsch, S.~Weidt, A.~G. Fowler, K.~Mølmer, S.~J. Devitt, C.~Wunderlich,
  and W.~K. Hensinger, ``Blueprint for a microwave trapped ion quantum
  computer,'' \emph{Sci. Adv.}, vol.~3, no. 2, e1601540, 2017.

\bibitem{rudolph2017optimistic}
T.~Rudolph, ``Why {I} am optimistic about the silicon-photonic route to quantum
  computing,'' \emph{APL Photonics}, vol.~2, no. 3, 030901, 2017.

\bibitem{yang2020operation}
C.~H. Yang \emph{et~al.}, ``Operation of a silicon quantum processor unit cell
  above one {Kelvin},'' \emph{Nature}, vol. 580, no. 7803, pp. 350--354, 2020.

\bibitem{ibmq}
``{IBM Quantum},'' 2022, https://quantum-computing.ibm.com/.

\bibitem{roffe2019quantum}
J.~Roffe, ``Quantum error correction: An introductory guide,'' \emph{Contemp.
  Phys.}, vol.~60, no.~3, pp. 226--245, 2019.

\bibitem{jahn2021holographic}
A.~Jahn and J.~Eisert, ``Holographic tensor network models and quantum error
  correction: A topical review,'' \emph{Quantum Sci. Technol.}, vol.~6, no. 3,
  033002, 2021.

\bibitem{lu2008experimental}
C.-Y. Lu, W.-B. Gao, J.~Zhang, X.-Q. Zhou, T.~Yang, and J.-W. Pan,
  ``Experimental quantum coding against qubit loss error,'' \emph{Proc. Natl.
  Acad. Sci.}, vol. 105, no.~32, pp. 11\,050--11\,054, 2008.

\bibitem{yang2010arbitrated}
Y.-G. Yang and Q.-Y. Wen, ``Arbitrated quantum signature of classical messages
  against collective amplitude damping noise,'' \emph{Opt. Commun.}, vol. 283,
  no.~16, pp. 3198--3201, 2010.

\bibitem{al2011reversing}
M.~A. Amri, M.~O. Scully, and M.~S. Zubairy, ``Reversing the weak measurement
  on a qubit,'' \emph{J Phys B At Mol Opt Phys.}, vol.~44, no. 16, 165509,
  2011.

\bibitem{fletcher2008channel}
A.~S. Fletcher, P.~W. Shor, and M.~Z. Win, ``Channel-adapted quantum error
  correction for the amplitude damping channel,'' \emph{IEEE Trans. Inf.
  Theory}, vol.~54, no.~12, pp. 5705--5718, 2008.

\bibitem{grassl2018quantum}
M.~Grassl, L.~Kong, Z.~Wei, Z.-Q. Yin, and B.~Zeng, ``Quantum error-correcting
  codes for qudit amplitude damping,'' \emph{IEEE Trans. Inf. Theory}, vol.~64,
  no.~6, pp. 4674--4685, 2018.

\bibitem{gong2022experimental}
M.~Gong, X.~Yuan, S.~Wang, Y.~Wu, Y.~Zhao, C.~Zha, S.~Li, Z.~Zhang, Q.~Zhao,
  Y.~Liu, F.~Liang, J.~Lin, Y.~Xu, H.~Deng, H.~Rong, H.~Lu, S.~C. Benjamin,
  C.-Z. Peng, X.~Ma \emph{et~al.}, ``Experimental exploration of five-qubit
  quantum error-correcting code with superconducting qubits,'' \emph{Natl. Sci.
  Rev.}, vol.~9, no. 1, nwab011, 2021.

\bibitem{superconduct}
S.~Markmann, C.~Reichl, W.~Wegscheider, and G.~Salis, ``Universal nuclear
  focusing of confined electron spins,'' \emph{Nat. Commun.}, vol.~10, no. 1,
  1097, 2019.

\bibitem{aleksandrowicz2019qiskit}
M.~S. Anis \emph{et~al.}, ``Qiskit: An open-source framework for quantum
  computing,'' 2021, {doi}: 10.5281/zenodo.2573505.

\bibitem{das2021design}
S.~Das, M.~S. Rahman, and M.~Majumdar, ``Design of a quantum repeater using
  quantum circuits and benchmarking its performance on an {IBM} quantum
  computer,'' \emph{Quantum Inf. Process.}, vol.~20, no. 7, 245, 2021.

\bibitem{zhukov2019quantum}
A.~A. Zhukov, E.~O. Kiktenko, A.~A. Elistratov, W.~V. Pogosov, and Y.~E.
  Lozovik, ``Quantum communication protocols as a benchmark for programmable
  quantum computers,'' \emph{Quantum Inf. Process.}, vol.~18, no. 1, 31, 2018.

\end{thebibliography}
\end{document}